# Evaluating Uncertainties in Electricity Markets via Machine Learning and Quantum Computing

Shuyang Zhu, *Student Member, IEEE*, Ziqing Zhu, *Member, IEEE*, Linghua Zhu, Yujian Ye, *Senior Member, IEEE*, Siqi Bu, *Senior Member, IEEE*, Sasa Z. Djokic, *Senior Member, IEEE*

*Abstract*— **The analysis of decision-making process in electricity markets is crucial for understanding and resolving issues related to market manipulation and reduced social welfare. Traditional Multi-Agent Reinforcement Learning (MARL) method can model decision-making of generation companies (GENCOs), but faces challenges due to uncertainties in policy functions, reward functions, and inter-agent interactions. Quantum computing offers a promising solution to resolve these uncertainties, and this paper introduces the Quantum Multi-Agent Deep Q-Network (Q-MADQN) method, which integrates variational quantum circuits into the traditional MARL framework. The main contributions of the paper are: identifying the correspondence between market uncertainties and quantum properties, proposing the Q-MADQN algorithm for simulating electricity market bidding, and demonstrating that Q-MADQN can capture a wider range of market equilibrium points, compared to conventional methods, without compromising computational efficiency. The proposed method is illustrated on IEEE 30-bus test network, confirming that it offers a more accurate model for simulating complex market dynamics.**

## I. INTRODUCTION

T HE evaluation of decision-making processes that are implemented by electricity market participants (e.g., generation companies, GENCOs) is very important for market operators, as it will allow them to observe and identify actions that may lead to imperfect competition, or to a reduced overall social welfare. Better understanding of decision-making will also enable more effective market optimization [1].

The use of traditional methods for modelling GENCOs' decision-making process, such as Multi-Agent Reinforcement Learning (MARL) [2]-[4], assumes that each GENCO is an independent smart agent, optimizing its bidding strategy with a *policy function* to maximize the *reward function*. However, this method faces three main challenges: 1) the GENCOs cannot establish a deterministic policy function due to uncertainties from long-term strategic plans, asset conditions, and traders' risk preferences and (ir)rationality; 2) the reward functions are uncertain and implicit, as GENCOs cannot observe complete information of the market, and therefore cannot predict the result of market clearing; 3) the reward of each GENCO is influenced by the decisions of other GENCOs, as they compete in market clearing.

This paper asserts that the properties of quantum computing are well-suited to improve simulation and evaluation of these three types of uncertainties. The presented results confirm that traditional MARL can be modified into quantum MARL, allowing to better simulate the equilibrium and dynamics of electricity markets. Specifically, the main contributions of this paper are as follows:

- Identification of a detailed correspondence between uncertainties in electricity market bidding game and properties of quantum computing.
- Proposal of a Quantum Multi-Agent Deep Q-Network algorithm (Q-MADQN) for simulating the bidding in electricity market, which uses variational quantum circuits to replace the traditional MADQN algorithm.
- Demonstration that the Q-MADQN algorithm can capture a wider variety of possible market equilibrium points, compared to traditional MADQN algorithms, while maintaining similar computational speeds.

The paper offers a feasible and potentially more accurate alternative for modeling complex market dynamics, and is organized as follows. Section II discusses the correspondence between uncertainties in electricity markets and properties of quantum computing, Section III gives details on the proposed method, Section IV introduces the case study and presents results of the analysis, and Section V lists main conclusions.

## II. CORRESPONDENCE BETWEEN UNCERTAINTIES IN ELECTRICITY MARKET BIDDING GAME AND PROPERTIES OF QUANTUM COMPUTING

This section provides some definitions of main terms and concepts in conventional MARL/MADQN and quantum computing methods, which are important for evaluation of the bidding game in a day-ahead electricity market.

**Definition 1:** *(Agent and Markov Decision Process, MDP)* An agent in MARL learns to act within an environment by interacting with it and receiving feedback. The environment is defined by the MDP, including *state* (observation), *action* (decision), *reward* (profit), and *state transition* after decision. Assuming that each GENCO acts as an agent, the state is the total demand in the day-ahead market for each hour (or half-hour), actions are the bids submitted by each GENCO for each hour, and reward is the revenue obtained after market clearing for each hour. The MDP is considered over a 24-hour horizon, with GENCOs optimizing their bids to maximize daily revenue.

**Definition 2:** *(Qubit, Superposition, and Entanglement)* A Quantum bit (qubit) is the fundamental unit of a quantum system, analogous to a bit in classical computing. In this paper, the qubits represent elements of the MDP. Unlike a classical bit, which can only be either 0 or 1, a qubit can be in a superposition state, representing both 0 and 1 simultaneously. Entanglement represents connections between quantum states, where state of one qubit affects the state of another, regardless of distance, or presence of any explicit connection between them.

Based on the above definitions, three properties of quantum computing are suitable for addressing challenges in Section I.

**Property 1:** *(To address Challenge 1)* Qubits representing actions can exist in superposition, enabling the modeling of GENCO's bidding decisions with unpredictable uncertainties due to non-perfect rationality and risk preferences. This differs from MARL, where a deterministic policy function is assumed.

**Property 2:** *(To address Challenge 2)* Qubits representing reward and next states can exist in superposition, modeling the GENCO agent's bidding results as a spectrum of possibilities before market clearing. Once market clearing is completed, the quantum state collapses to reveal the outcome.

**Property 3:** *(To address Challenge 3)* The paper defines two types of entanglement: generation and economic. Generation entanglement refers to the immediate balancing of power output among GENCOs, i.e., a change in one GENCO's output causes other to adjust their generation to meet the system demand. Economic entanglement builds on this, where the rewards of GENCOs are intertwined with their bidding strategies and power outputs. Thus, generation entanglement inherently leads to economic entanglement, as revenues are impacted by operational adjustments and market interactions.

## III. ALGORITHM DESIGN

### A. Overall Framework

This section introduces a general framework for integrating quantum computing into conventional MARL method, using the MADQN algorithm [5] as the baseline. In MADQN, each GENCO agent makes their own decision based on the estimated benefits of all possible actions using the Q-function, which is updated by:

$$Q_{new}(s,a) \leftarrow Q(s,a) + \alpha[r + \gamma \max_{a'} Q(s',a') - Q(s,a)] \quad (1)$$

where: $(s, a)$ is the estimated Q-value for a given state $s$ and action $a$, $\alpha$ is the learning rate, $r$ is the reward received after taking action $a$ in state $s$, $\gamma$ is the discount factor, $s'$ is the new state after action $a$ is taken, and $a'$ represents possible future actions from state $s'$.

In the MADQN, a neural network parameterized by $\theta$ is used to approximate the Q-function. It takes a state as the input and outputs Q-values for all possible actions. The parameters are updated using gradient descent method, and loss function is defined to reflect an expectation, is given by:

$$L(\theta) = E[(r + \gamma \max_{a'} Q_{\theta^-}(s',a') - Q_{\theta}(s,a))^2] \quad (2)$$

where: $\theta^-$ denotes the target network (differentiated from the current network $\theta$) designated to stabilize the training.

When the MDP elements (state, action, rewards) are represented by qubits, the gradient descent to update the Q-function can be implemented by:

$$\theta \leftarrow \theta + \alpha \cdot \nabla_\theta L(\langle O \rangle_{s,\theta}) \quad (3)$$

where: $\theta$ is the *variational* parameter in the quantum circuit, analogous to a classical parameter in conventional neural networks, $\alpha$ denotes the learning rate, and $L(\langle O \rangle_{s,\theta})$ is the loss function, in which the *variational* Q-function $\langle O \rangle_{s,\theta}$ is estimated by the *Variational Quantum Circuit* (VQC), with input $s$ and variational parameter $\theta$.

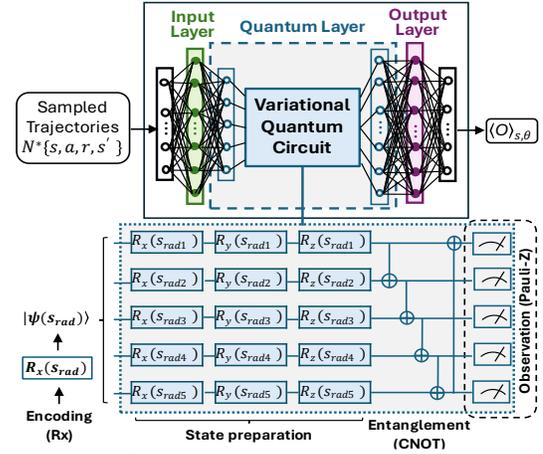

Fig. 1. Architecture of the VQC design.

### B. Details of the VQC Design

The network architecture, depicted in Fig.1, comprises three main layers: the input layer, the quantum layer, and the output layer. The input and output layers are classical fully connected layers, designated for receiving state $s$ and producing Q-function $\langle O \rangle_{s,\theta}$, respectively. The quantum layer features a quantum circuit enhanced by two additional fully connected layers, to align the dimensions between the VQC and the input and output layers. Main steps of operation of quantum circuit for enabling quantum properties introduced in Section II are detailed next.

**Step 1: Encoding** This step involves translating classical input data $s$ into quantum states before entering the VQC. First, $s$ is pre-normalized within $[0, \pi]$ and converted into radians, resulting in radial classical data $s_{rad}$. Then, the $R_x$ gate applies a rotation on $s_{rad}$ about the x-axis in the Bloch sphere (space of quantum systems), defined in (4). This rotation embeds the radial classical data $s_{rad}$ into the quantum states, $|\psi(s_{rad})\rangle$.

$$R_x(s_{rad}) = \begin{bmatrix} cos(s_{rad}/2) & -isin(s_{rad}/2) \\ -isin(s_{rad}/2) & cos(s_{rad}/2) \end{bmatrix} \quad (4)$$

**Step 2: Entanglement** The goal of this step is to generate correlations among qubits within the VQC. This is achieved by applying $R_x$, $R_y$, and $R_z$ gates for state preparation and CNOT gates for entanglement. Similar to $R_x$ in (4), $R_y$, and $R_z$ are defined as:

$$R_y(s_{rad}) = \begin{bmatrix} cos(s_{rad}/2) & -sin(s_{rad}/2) \\ sin(s_{rad}/2) & cos(s_{rad}/2) \end{bmatrix} \quad (5)$$

$$R_z(s_{rad}) = \begin{bmatrix} e^{-is_{rad}/2} & 0 \\ 0 & e^{is_{rad}/2} \end{bmatrix} \quad (6)$$

The CNOT gate is a two-qubit gate that flips the state of the target qubit if the control qubit is in the state $|1\rangle$, with its matrix representation in (7). The CNOT gates are arranged in a ladder topology between five qubits, indicated in Fig. 1, to ensure all qubits are entangled.

$$CNOT = \begin{bmatrix} 1 & 0 & 0 & 0 \\ 0 & 1 & 0 & 0 \\ 0 & 0 & 0 & 1 \\ 0 & 0 & 1 & 0 \end{bmatrix} \quad (7)$$

**Step 3: Observation** This step involves measuring quantum states to extract classical information using Pauli-Z operator, defined by the matrix $Z = \begin{bmatrix} 1 & 0 \\ 0 & -1 \end{bmatrix}$. The Pauli-Z operator has eigenvalues of +1 and -1, corresponding to the computational basis states $|0\rangle$ and $|1\rangle$. When measuring a qubit in the Pauli-Z basis, the quantum state collapses into one of basis states, providing the classical information needed for further processing. The observed results, $\langle\sigma_z\rangle$, are scaled back to the original classical data range before being input into following neural network. To optimize the entire network, the classical Adam optimizer is used to fine-tune the parameters via gradient descent shown in (3), with loss function formulated in (2).

## IV. CASE STUDY

A test market study was developed based on the IEEE 30-bus system with 6 GENCO agents with specific generation costs in [6] for comparison of the proposed Q-MADQN and conventional MADQN models. The hourly load demand profile is also extracted from [6], and a price cap of [0, 1000] USD/MWh is set. The output are the converged bidding strategies of each GENCO agent, i.e., the market equilibrium point. The following early stopping criteria is used to identify the market equilibrium points (i.e., the algorithm is converged): the total daily reward has a lower than a 5% change over the five consecutive episodes. MADQN converged in 483 episodes, while Q-MADQN required 788 episodes. Due to the parallelism of computation enabled by quantum superposition, the computational time of each invocation of VQC for exploration is comparable to that of MADQN calling the deep network, approximately 0.01 seconds.

Fig. 2 and 3 show distributions of state-action pair and state-reward pair of GENCO 4 (with highest marginal fuel cost) of MADQN and Q-MADQN at 18:00 PM, with point size and color indicating frequency, demonstrating Q-MADQN explores the strategy space more extensively and also simulates more high-bid actions than MADQN, leading to higher rewards, as its distributions are more dispersed. Table I records market marginal cost at two selected times (valley & peak) and total daily reward of six GENCOs under strategic bids ($MC_S$, $R_S$) and actual-cost bids ($MC_A$, $R_A$) scenarios. Q-MADQN reveals a potential behavior that traditional MADQN fails to detect, leading to reduced social welfare. Firstly, market equilibrium points obtained by Q-MADQN yield much greater total gains for GENCOs compared to actual-cost bids. Secondly, Q-MADQN results in an excessively high marginal cost at peak demand, as all GENCOs are guaranteed dispatch, and this enables them to engage in strategic bidding by approaching price cap, thus abusing market power. Additionally, Q-

MADQN intelligently simulates some GENCOs reducing bids to secure dispatch during off-peak periods.

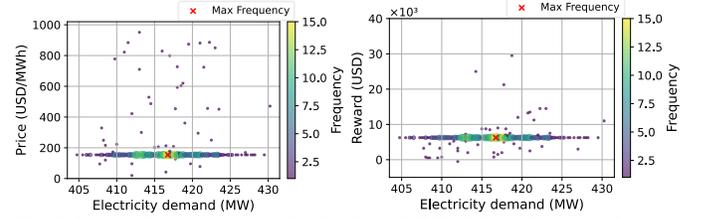

Fig. 2. State-action and reward distributions of GENCO 5 using MADQN.

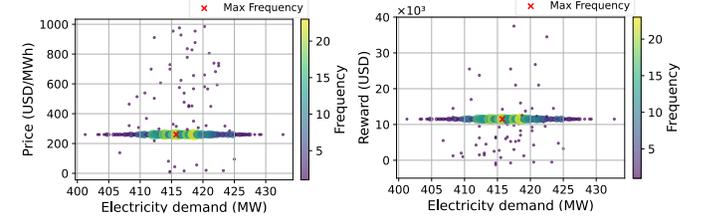

Fig. 3. State-action and reward distributions of GENCO 5 using Q-MADQN.

TABLE I. COMPARISON OF TWO ALGORITHMS UNDER TWO SCENARIOS.

|  |  | Strategic Bids | | Actual-Cost Bids | |
|---|---|---|---|---|---|
|  |  | $MC_S$ | $R_S$ | $MC_A$ | $R_A$ |
| MADQN | 06:00 | 154 | 880101 | 10 | -11234 |
|  | 18:00 | 726 |  | 24 |  |
| Q-MADQN | 06:00 | 260 | 1465441 | 10 | -11624 |
|  | 18:00 | 986 |  | 24 |  |

## V. CONCLUSIONS

This paper introduces Q-MADQN algorithm, which integrates variational quantum circuits into the MARL framework, to effectively evaluate and simulate uncertainties in the electricity market. The results confirm that Q-MADQN allows for a more thorough exploration of the entire state-action space and simulates more potential bidding strategies of profit-oriented GENCOs that could increase social welfare loss than MADQN, providing valuable insights for market designers. Future research will tackle the challenges of large-scale market simulations and real-world electricity market data will be used to refine Q-MADQN's effectiveness in practical applications.

## REFERENCES

[1] G. Mohy-ud-din, K. M. Muttaqi, and D. Sutanto, "A Cooperative Energy Transaction Model for VPP Integrated Renewable Energy Hubs in Deregulated Electricity Markets," *IEEE Trans. Ind. Appl.*, vol. 58, no. 6, pp. 1–14, 2022.

[2] Y. Ye, Y. Tang, H. Wang, X.-P. Zhang, and G. Strbac, "A Scalable Privacy-Preserving Multi-Agent Deep Reinforcement Learning Approach for Large-Scale Peer-to-Peer Transactive Energy Trading," *IEEE Trans. Smart Grid*, vol. 12, no. 6, pp. 5185–5200, 2021.

[3] I. N. Moghaddam, B. Chowdhury, and M. Doostan, "Optimal Sizing and Operation of Battery Energy Storage Systems Connected to Wind Farms Participating in Electricity Markets," *IEEE Trans. Sustain. Energy*, vol. 10, no. 3, pp. 1184–1193, 2019.

[4] Y. Ye, D. Papadaskalopoulos, J. Kazempour, and G. Strbac, "Incorporating Non-Convex Operating Characteristics Into Bi-Level Optimization Electricity Market Models," *IEEE Trans. Power Syst.*, vol. 35, no. 1, pp. 163–176, 2020.

[5] D. Qiu, Y. Ye, D. Papadaskalopoulos, and G. Strbac, "Scalable coordinated management of peer-to-peer energy trading: A multi-cluster deep reinforcement learning approach," Appl. Energy, vol. 292, pp. 116940–, 2021.

[6] Q. Duan, J. Liu, K. Zeng, J. Zhong and Y . Guo, "Integrated scheduling of generation and demand shifting in day-ahead electricity market," *Int. Trans. Electr. Energy Syst.*, vol. 29, no. 7, e2843, 2019.